\documentclass[aps,pra,reprint,amsmath,floatfix,amssymb,superscriptaddress]{revtex4-2}
\usepackage{graphicx}
\usepackage{physics}
\usepackage{dcolumn}
\usepackage{bm}
\usepackage{color}
\usepackage{hyperref}
\hypersetup{colorlinks,allcolors=blue}
\usepackage{orcidlink}
\newcommand{\be}{\begin{equation}}
\newcommand{\ee}{\end{equation}}
\newcommand{\bea}{\begin{eqnarray}}
\newcommand{\eea}{\end{eqnarray}}
\newcommand{\ba}[1]{\begin{array}{#1}}
\newcommand{\ea}{\end{array}}

\begin{document}

\title{Collective-State Preparation in a Subwavelength Triangular Trimer Using SUPER Excitation}

\author{Thomas Nunner\,\orcidlink{0009-0001-7853-2635}}
\affiliation{Institut für Theoretische Physik, Universität Innsbruck, Technikerstraße 21a, A-6020 Innsbruck, Austria}
\author{Johannes Kerber\,\orcidlink{0009-0002-1957-8008}}
\affiliation{Institut für Theoretische Physik, Universität Innsbruck, Technikerstraße 21a, A-6020 Innsbruck, Austria}
\author{Helmut Ritsch\,\orcidlink{0000-0001-7013-5208}}
\affiliation{Institut für Theoretische Physik, Universität Innsbruck, Technikerstraße 21a, A-6020 Innsbruck, Austria}
\author{Arpita Pal\,\orcidlink{0000-0003-1124-6818}}
\email{Arpita.Pal@uibk.ac.at}
\affiliation{Institut für Theoretische Physik, Universität Innsbruck, Technikerstraße 21a, A-6020 Innsbruck, Austria}

\date{\today}

\begin{abstract}
The Swing-UP of quantum EmitteR population (SUPER) scheme has recently been proposed as a deterministic method for the preparation of collective radiative states in two strongly dipole-coupled quantum emitters (Phys. Rev. Res. \textbf{8}, 013179 (2026)). Here, we apply this approach to an equilateral subwavelength triangular trimer of dipole-coupled two-level quantum emitters (QEs), loosely inspired by biological light-harvesting ring geometries, and demonstrate that tailored, time-overlapping, red-detuned ultrashort SUPER pulses can selectively prepare collective states in a $C_3$-symmetric system, including energetically degenerate eigenstates that are resolved through site-dependent optical phases. We find that both the state selectivity and the preparation efficiency depend strongly on the inter-emitter spacing. In particular, at deep-subwavelength separations, the symmetric collective state can be deterministically prepared with near-unity efficiency (approximately $94\%$), whereas the inversion efficiency and state selectivity are significantly lower at larger inter-emitter separations. Furthermore, this state preparation technique inherits a certain degree of robustness against reasonable static position imperfections and on-site frequency inhomogeneities of the individual QEs. Our results demonstrate that deep-subwavelength triangular trimers and, more broadly, highly compact ring geometries are excellent candidates for the deterministic preparation of collective radiative states via SUPER excitation. These predictions could be realized with solid-state emitters and molecules. Our findings offer a route toward the direct probing of the `pure' electromagnetic layer of interaction in biological and bio-inspired synthetic nanophotonic ring configurations, with possible relevance in photonics, quantum information processing, and metrology.
\end{abstract}

\maketitle

\section{Introduction}
Natural light-harvesting (LH) complexes in purple photosynthetic bacteria exhibit ordered circular assemblies of pigments embedded in protein scaffolds~\cite{mcdermott:nature:1995, kuhlbrandt:structure:1995}, where geometry plays a crucial role in efficient light absorption, excitation delocalization, and near lossless energy transport toward the reaction center~\cite{fleming:arpc:2009, scholes:cr:2017, mennucci:rmp:2018, kohler_2006}. The extraordinary efficiency of biological LH systems has fascinated researchers across chemistry, biology, and physics for decades~\cite{VANGRONDELLE19941, Grondelle:pccp:2006, Ishizaki:pccp:2010, scholes:arpc:2015, mennucci:cr:2017}, motivating extensive research efforts to understand the underlying physical principles governing light capture and energy transport. Such insights contribute towards the development of artificial light-harvesting materials, molecular aggregates, quantum nanophotonic devices, and energy transport technologies that inherit the efficiency of biological LH systems~\cite{patra:cr:2017, scholes:cr:2017}.

Taking inspiration from these biological ring architectures, subwavelength-spaced circular assemblies of quantum emitters (QEs), coupled via dipole-dipole interactions~\cite{FICEK:PR:2002}, have emerged as a promising model platform in quantum optics and nanophotonics. Theoretical quantum optical studies illustrate that ring geometries can facilitate collective mechanisms such as dark-state-mediated excitation transfer~\cite{moreno:pra:2019, Needham_2019} and function as highly efficient optical antennas with enhanced light-capture capabilities~\cite{Mattiotti:njp:2021, Moreno-Cardoner:oe:22, erik:prxe:2023, Erik:prl:2025}. Also, collective quantum optical treatment of the biological stacked LH2 ring geometry could provide a natural blueprint for designing nanoscale layered ring architectures that can support highly efficient and asymmetric inter-layer excitation energy transfer mechanism~\cite{pal:njp:2025}. However, the validity of collective optical phenomena~\cite{Dicke:pr:1954,GROSS:pr:1982} in natural LH systems is not fully established yet~\cite{pal:njp:2025, pal:arxiv:2026}. Real biological complexes operate at room temperature in wet and messy environments, where vibronic coupling, disorder, and nontrivial system-environment interactions play an essential role for its overall functionality~\cite{Ishizaki:pccp:2010, Fuller:nc:2014, Fassioli:jrsi:2014, mennucci:rmp:2018}. Nevertheless, relying on a generic simple collective quantum optical description offers opportunities for partially isolating the `pure' electromagnetic layer of interaction and thereby helps to identify a possible importance of the geometric aspect of biological architectures in its relevant optical functionality~\cite{pal:njp:2025, pal:arxiv:2026}.

Subwavelength-spaced QE assemblies are a natural platform to explore collective light-matter interactions~\cite{genes:prxq:2022, janne:pra:2023}. When emitters are separated by distances smaller than the resonant wavelength, strong dipole-dipole interactions generate eigenstates with modified radiative properties that differ strongly from those of the independent emitters~\cite{FICEK:PR:2002, brandes:pr:2005}, exhibiting enhanced (superradiance) or suppressed decay (subradiance). Such collective radiative states are interesting for a broad range of investigations in nanophotonics and cooperative quantum optics~\cite{brandes:pr:2005, kaiser:prl:2012, genes:prxq:2022, tiranov:science:2023, janne:pra:2023}. Very recently it has been illustrated that collective effects in aspects of light-harvesting, in particular, emission, absorption, and transfer could be understood within a common framework ~\cite{kushwaha:arxiv:2025}, further emphasizing the importance of cooperative quantum optics.

A key challenge at this research frontier is selective and robust preparation of specific collective radiative states of interacting QEs~\cite{genes:sr:2015} amid a dissipative environment. Recently, the Swing-UP of quantum EmitteR population (SUPER) scheme was introduced as a pulse-based excitation mechanism for efficient preparation of a specific excited state in driven single quantum dot (QD) system~\cite{doris:prxq:2021, Karli:nl:2022}. More recently, it has been shown that the SUPER scheme allows one to selectively populate collective states in two strongly coupled, deep-subwavelength-spaced QEs, providing access to their distinct radiative features amid reasonable environmental decoherence~\cite{kerber:prr:2026}. While collective eigenstates and radiative decay behavior in triangular geometries had been studied before~\cite{Ostermann:oe:12, zhu:pra:2013, Linsa:arxiv:2026}, the SUPER excitation has not been applied to closed QE geometries, to selectively prepare collective radiative states. The geometry plays a governing role in determining the energy levels of the decay cascade. The equilateral triangular trimer possesses a qualitatively different level structure from both the coupled two-emitter case~\cite{kerber:prr:2026} and a linear three-emitter chain~\cite{Ostermann:oe:12}. In particular, the $C_3$ rotational symmetry of the triangle introduces degenerate pairs of dressed states, a feature that has no counterpart in either the dimer or the linear trimer. Selectively preparing a chosen state within such a degenerate manifold is distinct in the case of closed very compact ring geometries.

In this work, we utilize the SUPER excitation scheme for the generic case of an equilateral triangular trimer of three dipole-coupled, two-level QEs in free space, for populating certain collective states. Our numerical investigation shows that the efficiency of selective collective-state preparation depends strongly on geometric compactness. The deep-subwavelength regime helps for near-unity inversion into specific collective states. As the inter-emitter separation increases, the collective energy shifts decrease, the collective decay rates are modified, and preparation efficiency into a certain target collective state becomes smaller. We further discuss the robustness of this population inversion mechanism against reasonably small on-site static position and frequency disorders, inheriting the very own signature of the SUPER excitation scheme.

The manuscript is organized as follows. Section~\ref{theory} details the theoretical framework of our investigation (together with Appendix~\ref{green}). Section~\ref{selective preparation} presents our findings on the selective preparation of a target collective radiative state in a subwavelength trimer geometry (together with Appendix~\ref{b1-pop}) and also provides qualitative insights into the relevance of our investigation for biological LH ring geometries. In Section~\ref{disorder} we provide quantification of the robustness of the prepared collective target state against environmental decoherence, i.e., against static on-site position and frequency disorders. Section~\ref{expt} discusses experimental platforms for our proposal, and finally in Section~\ref{conclusions} we present conclusions and outlook of our study.

\begin{figure}[b]
\centering
\includegraphics[width=0.9\linewidth]{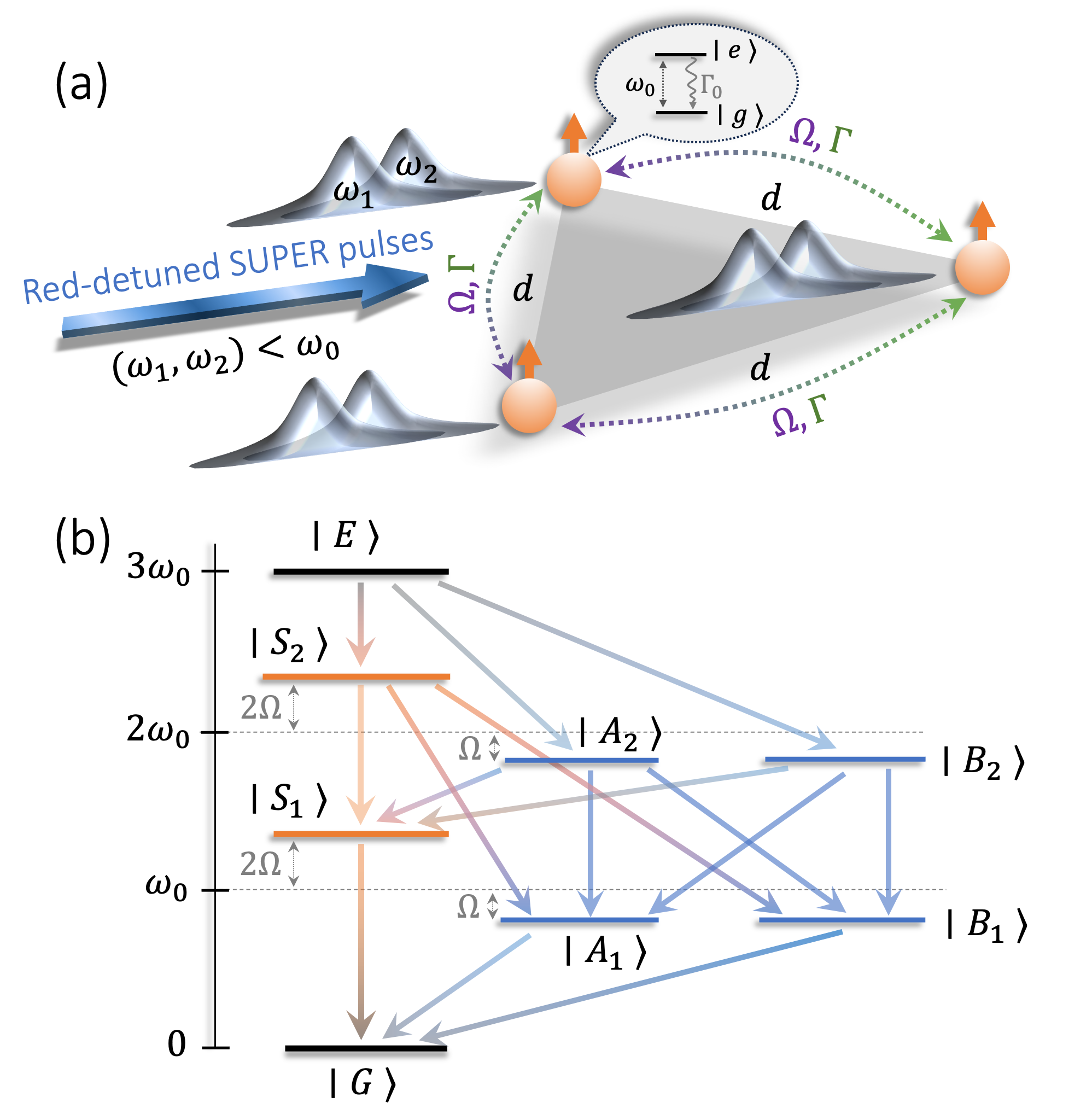}
\caption{(a) Schematic of an equilateral triangular arrangement of three two-level QEs separated by a distance $d$ and driven by parameter-optimized sets of SUPER pulses. Each QE has a ground state $|g\rangle$ and an excited state $|e\rangle$, with transition frequency $\omega_0$ (corresponding resonant transition wavelength is $\lambda_0$) and spontaneous emission rate $\Gamma_0$. The SUPER pulses are red-detuned with frequencies $(\omega_1, \omega_2) < \omega_0$. Dipole-dipole interactions between the QEs induce collective energy shifts $\Omega$ and decay rates $\Gamma$ depending on the inter-emitter separation and the dipole orientations (see text). The deep-subwavelength regime ($d \ll \lambda_0$) enables very high fidelity population inversion into a target collective state. (b) Possible decay channels for the dipole-coupled equilateral triangular trimer. Downward arrows indicate the allowed radiative decay pathways between the symmetry-adapted dressed eigenstates. The explicit forms of the eigenstates and their energies are listed in Table~\ref{tab:eigen}.}
\label{schematic}
\end{figure}

\section{Theoretical Framework}
\label{theory}

We consider an equilateral triangle of three QEs ($N=3$) as a simplest closed emitter (ring) arrangement, loosely inspired by biological LH2 complexes. The three identical two-level QEs are coupled via dipole-dipole interaction and driven by a pulsed excitation field $\Omega(t)$ (see Fig.~\ref{schematic}(a)). Setting $\hbar \equiv 1$ the corresponding Hamiltonian reads as follows

\begin{align}
    \hat{H}(t) = \omega_0\sum_{i}\hat{\sigma}^+_i\hat{\sigma}^-_i 
    +& \sum_{i\neq j}\Omega_{ij}\hat{\sigma}^+_i\hat{\sigma}^-_j\nonumber\\
    -& \frac{1}{2}\sum_{i}(\Omega(t)\hat{\sigma}^+_i + \mathrm{h.c.}).
    \label{eq:ham}
\end{align}
The three terms represent, respectively, the
on-site energy, the dipole-dipole coupling, and the pulsed drive, where $i,j\in\{1,2,3\}$.  Transition frequency of each QE is $\omega_0$, $\hat{\sigma}_i^{\pm}$ are the raising and lowering operators of the $i^{\text{th}}$ QE, and $\Omega_{ij}$ is the distance- and orientation-dependent collective energy shift between the $i^{\text{th}}$ and $j^{\text{th}}$ QE (see Appendix~\ref{green}). For weak system-environment interaction the system dynamics obeys the Lindblad master equation
\begin{align}
    \partial_t\hat{\rho}(t) = -i[\hat{H}(t),\hat{\rho}(t)] + 
    \hat{\mathcal{L}}[\hat{\rho}(t)]~,\label{eq:master}
\end{align}
with the Liouvillian superoperator
\begin{align}
    \hat{\mathcal{L}}[\hat{\rho}] = \sum_{ij}\frac{\Gamma_{ij}}{2}
    \left(2\hat{\sigma}^-_i\hat{\rho}\hat{\sigma}^+_j
    -\hat{\sigma}^+_j\hat{\sigma}^-_i\hat{\rho} 
    - \hat{\rho}\hat{\sigma}^+_j\hat{\sigma}^-_i\right)~,
\end{align}
where $\hat{\rho}$ is the density matrix and $\Gamma_{ij}$ are the collective decay rates. See Appendix~\ref{green} for further details.

\subsection{SUPER Excitation}
A SUPER excitation pulse can be expressed as~\cite{doris:prxq:2021}
\begin{align}
    \Omega_{\text{S}}(t) = \Omega^1_{\text{S}}(t)e^{-i\omega_1 t} + 
    \Omega^2_{\text{S}}(t-\tau)e^{-i\omega_2 t}~,
    \label{eq:SUPER_pulse}
\end{align}
consisting of two temporally overlapping Gaussian envelopes $\Omega^i_{\text{S}}(t) = \left(\alpha_i/\sqrt{2\pi\sigma_i^2}\right)\exp\{-{t^2}/{(2\sigma_i^2)}\}$ with pulse areas $\alpha_i$, temporal widths $\sigma_i$ and time shift between the pulses of $\tau$. In a rotating frame at frequency $\omega_0$ the Hamiltonian in Eq.~(\ref{eq:ham}) becomes
\begin{align}
    \hat{H}_{\text{S}}(t) =  \sum_{i\neq j}\Omega_{ij}\hat{\sigma}^+_i\hat{\sigma}^-_j
    - \frac{1}{2}\sum_{j}\left( \tilde{\Omega}_{\text{S}}(t)\hat{\sigma}^+_j
    e^{i\vartheta_j} + \mathrm{h.c.}\right)~, \label{eq:ham_tra}
\end{align}
with modified form of SUPER pulses which is
\begin{align}
    \tilde{\Omega}_{\text{S}}(t) = \Omega^1_{\text{S}}(t)e^{-i\Delta_1 t} + 
    \Omega^2_{\text{S}}(t-\tau)e^{-i\Delta_2 t}~,\label{eq:SUPER_transf}
\end{align}
and detunings $\Delta_i = \omega_i - \omega_0$. The site-dependent tunable relative optical phases $\vartheta_j$s (in Eq.~(\ref{eq:ham_tra})) allow the SUPER sets of pulses to selectively address specific collective states~\cite{kerber:prr:2026}. 
We solve the master equation in Eq.~(\ref{eq:master}) using the Hamiltonian of Eq.~(\ref{eq:ham_tra}) to obtain the results.

\subsection{Collective Eigenstates and Decay Cascade}
The Hamiltonian for the non-driven dipole-coupled triangular trimer can be expressed as
\begin{align}
    \hat{H}_{\text{DD}} =\omega_0\sum_{i}\hat{\sigma}^+_i\hat{\sigma}^-_i 
    + \Omega\sum_{i\neq j}\hat{\sigma}^+_i\hat{\sigma}^-_j~,
    \label{eq:ham_dd}
\end{align}
where the three-fold symmetry of the equilateral triangle leads to equal pairwise coupling, $\Omega_{ij} = \Omega$ for all $i \neq j$. The eight bare product states are $|eee\rangle$, $|eeg\rangle$, $|ege\rangle$, 
$|gee\rangle$, $|egg\rangle$, $|geg\rangle$, $|gge\rangle$, and $|ggg\rangle$. Diagonalizing $\hat{H}_{\text{DD}}$ yields the dressed eigenstates and eigenenergies, as listed in Table~\ref{tab:eigen}~\cite{Ostermann:oe:12}, and the resulting decay cascade is illustrated in Fig.~\ref{schematic}(b).

\begin{table}[h]
\caption{\label{tab:eigen} The eigenenergies and the dressed basis of the decay cascade of an equilateral triangle trimer interacting via dipole-dipole interaction~\cite{Ostermann:oe:12}. Corresponding level diagram is in Fig.~\ref{schematic}(b).}
\begin{ruledtabular}
\begin{tabular}{ll}
$E_j$ & $\ket{\text{Dressed~basis}}$ \\
\hline
$3\omega_0$           & $\ket{E}   = \ket{eee}$    \\
$2\omega_0+2\Omega$   & $\ket{S_2} = \frac{1}{\sqrt{3}}(\ket{eeg} + \ket{ege} + \ket{gee})$  \\
$2\omega_0 -\Omega$  & $\ket{B_2} = \frac{1}{\sqrt{2}}(\ket{eeg} - \ket{gee})$                \\
$2\omega_0 -\Omega$  & $\ket{A_2} = \frac{1}{\sqrt{6}}(\ket{eeg} - 2\ket{ege} + \ket{gee})$ \\
$\omega_0+2\Omega$    & $\ket{S_1} = \frac{1}{\sqrt{3}}(\ket{egg} + \ket{geg} + \ket{gge})$  \\
$\omega_0 -\Omega$   & $\ket{B_1} = \frac{1}{\sqrt{2}}(\ket{egg} - \ket{gge})$                \\
$\omega_0 -\Omega$   & $\ket{A_1} = \frac{1}{\sqrt{6}}(\ket{egg} - 2\ket{geg} + \ket{gge})$ \\
$0$                  & $\ket{G}   = \ket{ggg}$                                                 \\
\end{tabular}
\end{ruledtabular}
\end{table}

\section{Selective Preparation of Collective Radiative States}
\label{selective preparation}
\subsection{Inversion Efficiency and Geometric Compactness}
Let us now investigate the possibility of selective preparation of collective states in the triangular trimer, either in free space or in an environment with weak or controllable system-environment interaction where the coupled-dipole model is valid~\cite{pal:njp:2025}. By tuning the pulse parameters we selectively target states in the first-excitation manifold of the decay cascade (see Fig.~\ref{schematic}(b)). Note that one could in principle choose any excited collective state from the decay cascade.

\begin{figure}[h]
\centering
\includegraphics[width=\linewidth]{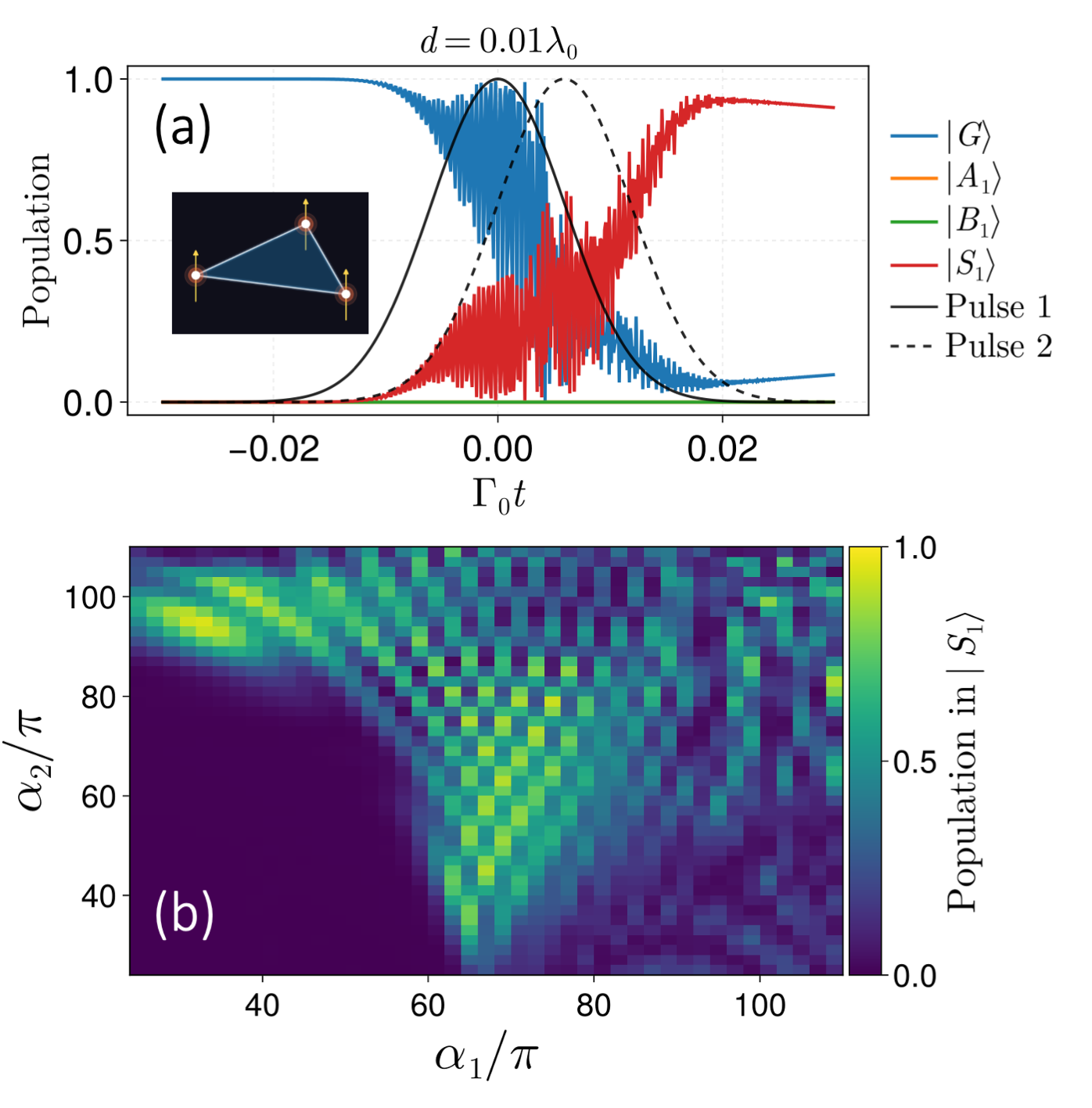}
\caption{Near-unity population inversion into the symmetric collective state $|S_1\rangle$ at deep-subwavelength QE separation $d = 0.01\lambda_0$ for transversely oriented dipoles (see inset). Panel (a) shows the time evolution of state populations under the SUPER excitation, with optimized pulse parameters $(\Delta_1, \Delta_2) = (-3.00, -12.75)$~meV, $(\alpha_1, \alpha_2) = (33\pi, 95\pi)$, $(\sigma_1, \sigma_2) = (6, 6)$~ps, $\vartheta_j = 0$ ($j = 1,2,3$), and $\tau = 5.9$~ps, for transversely oriented dipoles at nearest-neighbor separation $d = 0.01\lambda_0$. A final population of $94.3\%$ (approximately) is achieved in the symmetric state $|S_1\rangle$. The two pulse normalized envelopes are shown as black solid and dashed curves. Panel (b) shows the corresponding heatmap of the final $|S_1\rangle$ population as a function of pulse areas $(\alpha_1, \alpha_2)$, with all other parameters fixed at the values given in panel (a).}
\label{s1-inversion}
\end{figure}
\begin{figure}[h]
\centering
\includegraphics[width=\linewidth]{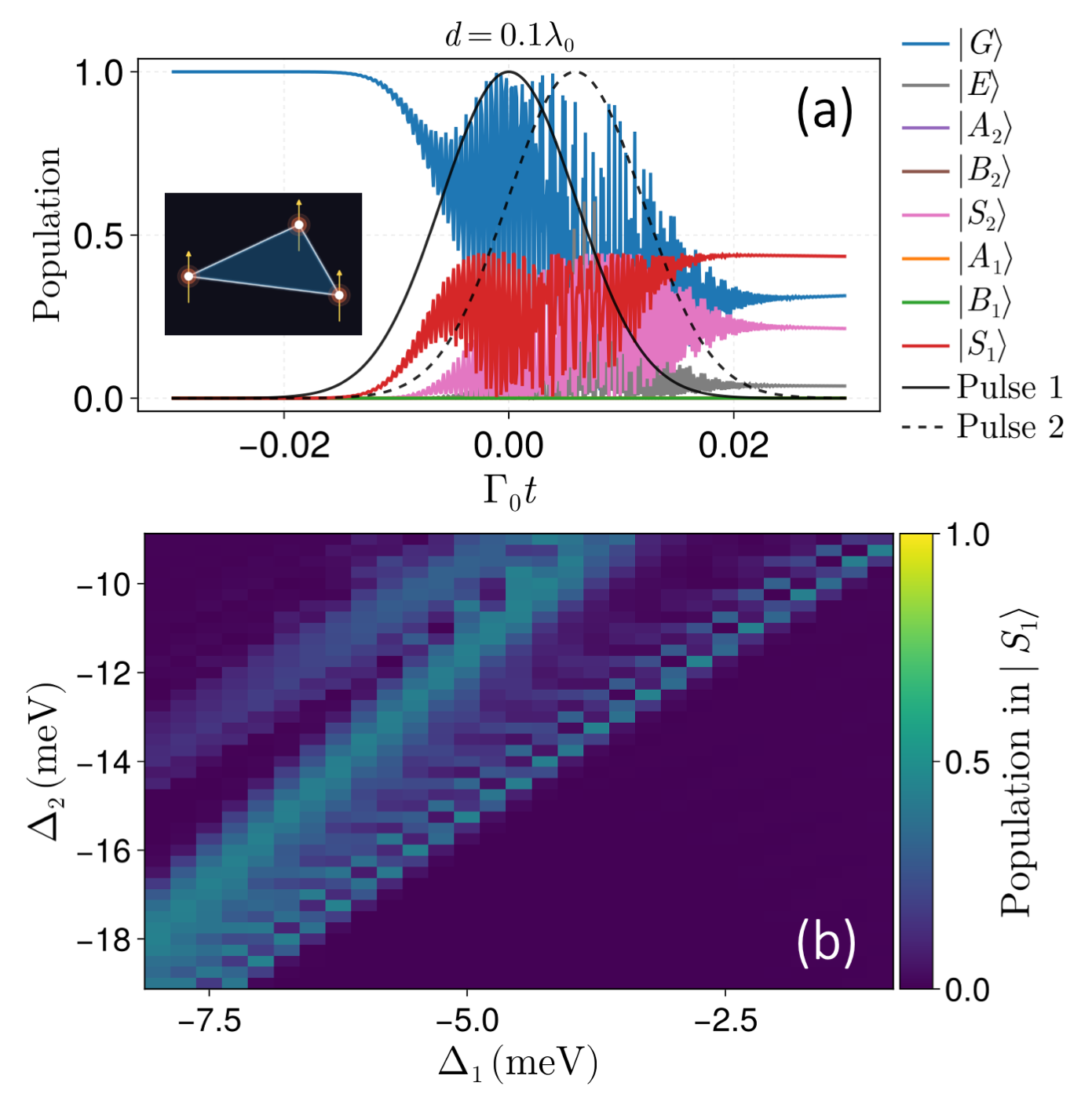}
\caption{Reduced selective collective-state preparation at larger inter-emitter separation $d = 0.1\lambda_0$ for transversely oriented dipoles (see inset). Panel (a) shows the time evolution of state populations under SUPER excitation for transversely oriented dipoles at inter-emitter separation $d = 0.1\lambda_0$, with pulse parameters $(\alpha_1, \alpha_2) = (33\pi, 95\pi)$, $\tau = 5.9$~ps, $(\Delta_1, \Delta_2) = (-3.25, -12.50)$~meV, $(\sigma_1, \sigma_2) = (6, 6)$~ps, and $\vartheta_j = 0$ ($j = 1, 2, 3$). The maximum achievable population in $|S_1\rangle$ is approximately $44\%$, with residual population distributed among several other states as indicated. Panel (b) shows the corresponding heatmap of the final $|S_1\rangle$ population as a function of detuning $(\Delta_1, \Delta_2)$, with all other parameters fixed at the values given in panel (a).}
\label{d0.1-inversion}
\end{figure}

At the deep-subwavelength separation, i.e., $d = 0.01\lambda_0$ and for transversely oriented dipoles (the angle of inclination of the dipole to the plane of the geometry is $\theta = 90^{\circ}$) the inversion efficiency into the symmetric state $|S_1\rangle$ (see Table~\ref{tab:eigen}) reaches approximately $94.3\%$ (Fig.~\ref{s1-inversion}(a)), and the corresponding heatmap over pulse areas $(\alpha_1, \alpha_2)$ is shown in Fig.~\ref{s1-inversion}(b). This indicates that multiple parameter domains yield near-unity population transfer efficiency. Increasing the nearest-neighbor separation to $d = 0.1\lambda_0$, the maximum inversion efficiency into state $|S_1\rangle$ drops to approximately $44\%$, with residual population distributed among several other collective states and ground state, as indicated in Fig.~\ref{d0.1-inversion}(a). This behavior is governed by the distance dependence of the collective energy shift $\Omega$, which depends on the inter-emitter separation $d$ and the dipole orientations, and oscillates as a function of $d$~\cite{FICEK:PR:2002}. In particular, it diverges as $d \rightarrow 0$ and vanishes at large inter-emitter separations. At $d = 0.01\lambda_0$, the collective energy shifts are large, so the dressed states are well-separated in the decay cascade (see Fig.~\ref{schematic}(b)), enabling the SUPER scheme to address a specific collective state and prepare with near-unity efficiency. With suitable choices of $\vartheta_j$'s one could also populate $\ket{B_1}$ very efficiently, see Appendix~\ref{b1-pop} for details. It is worth emphasizing that the states $\ket{B_1}$ and $\ket{A_1}$ are energetically degenerate (see Table~\ref{tab:eigen}). The ability to selectively address either state within this degenerate subspace is a distinct feature of the triangular trimer that is absent in the two-emitter case~\cite{kerber:prr:2026}, where the corresponding eigenstates are non-degenerate. In practice, the site-dependent relative optical phases $\vartheta_j$ are first chosen and fixed. The remaining pulse parameters (pulse areas $\alpha_i$, detunings $\Delta_i$, etc.) are then scanned to maximize the inversion efficiency. In this way, the controllable phase of the driving optical field provides a handle to distinguish excitation pathways that are otherwise indistinguishable in energy, enabling state-selective control within the degenerate manifold, at the limit of highly compact trimer geometry.

In contrast, at larger inter-emitter separation, i.e., $d = 0.1\lambda_0$, the collective energy shifts ($\Omega$) are strongly suppressed and the dressed-states become nearly degenerate. In this case the modified collective decay behavior, over the characteristic timescale of excitation, does not allow for efficient preparation of the target states during this pulsed excitation mechanism. As a result, the population gets redistributed across multiple states, which reflects a significant lowering of the efficiency of selective preparation of target states. The heatmap in Fig.~\ref{d0.1-inversion}(b) indicates that no suitable combination of detuning parameters $(\Delta_1, \Delta_2)$ yields high inversion efficiency at this larger inter-emitter separation. Similarly, a scan over pulse areas $(\alpha_1, \alpha_2)$ would also lead to the same conclusion. For completeness, we would like to note that for the triangular trimer these observations are expected to remain valid for any choice of dipole orientation of QEs (at both inter-emitter separations). Here we choose to populate state $\ket{S_1}$, however, as stated above, one could choose any desired excited state from the decay cascade and would obtain similar observations. For this, one needs to choose $\vartheta_j$ suitably~\cite{kerber:prr:2026}. 

It is also important to note that the pulse parameters we discuss above are a result of manual parameter tuning and thus represent only certain optimized sets. In reality, the optimization process would offer a very broad spectrum of allowed parameter regimes, and our results represent only a subset from those optimum parameter domains. In principle, it is therefore always possible to choose a suitable set for its realistic future implementation.

\subsection{Relevance to Biological LH Ring Geometries}

Biological LH ring geometries are extremely compact. For example, the LH2 rings with ninefold symmetry have a diameter of approximately $60$~\AA~\cite{mcdermott:nature:1995} with nearest-neighbor separation of the order $0.001\lambda_0$. Here $\lambda_0 \approx$ 800 or 850~nm corresponds to the $Q_y$ absorption band of bacteriochlorophyll. The relevant optical transition to the first excited state of the bacteriochlorophyll molecule (the $Q_y$ dipole) is weakly driven by sunlight so that we can generally restrict ourselves to the single excitation manifold as discussed in the manuscript.

Ref.~\cite{Cremer:njp:2020} illustrates that for extremely small ($d\ll 0.01\,\lambda_0$) and densely packed ring geometries ($N \gg 3$), the collective energy shifts of all modes simultaneously vanish at a dipole orientation approaching the well-known magic angle of $54.7^{\circ}$, an angle at which the interaction between two point dipoles is traditionally known to vanish. However, for a closed ring geometry at finite $N$, the collective energy shifts do not cancel automatically, i.e., the cancellation is non-trivial and reflects vanishing energy shifts at the magic angle of $54.7^{\circ}$ only in the dense-ring limit ($N\to\infty$,\,$R/\lambda_0\ll 1$). As noted above, at or in close vicinity of this dipole orientation selective preparation of individual collective states becomes difficult, since the dressed states are degenerate or nearly degenerate in the decay cascade. However, Ref.~\cite{pal:njp:2025} illustrates that for a ninefold ring, certain collective states in the first-excitation manifold reflect large collective energy shifts, making them in principle accessible with high efficiency via the SUPER excitation. However, certain collective states that remain closely spaced in the eigen-energy space, would be expected to be difficult to address individually.

Nevertheless, it should be noted, that the generic coupled-dipole description employed here provides only a very simple and partial picture of light harvesting and energy transport mechanisms in the biological LH complexes~\cite{pal:njp:2025}, as in reality the system-environment interaction is significantly more complex. Ref.~\cite{pal:arxiv:2026} shows that certain specific collective eigenmode can also account for the absence of sub-sevenfold symmetries in LH complexes. Continuing on this note, Ref.~\cite{pal:njp:2025} illustrates that the collective energy shift at certain modes could be large, suggesting that some collective radiative state could, in principle, be accessed with high efficiency using the SUPER excitation, if the system-environment interaction is reasonably weak. Future experiments could thereby directly test how much of this target collective radiative state preparation efficiency is realizable in practice. In doing so, such experiments would assess the validity of these relatively simple and tractable coupled-dipole models. To be precise, a quantitative treatment of these biological systems would, however, require extensive overhaul of the theoretical framework used in the present work, and looking at the complexity for now we leave this as our future research direction.

\section{Robustness Against Environmental Decoherence}
\label{disorder}
\begin{figure}[b]
\centering
\includegraphics[width=\linewidth]{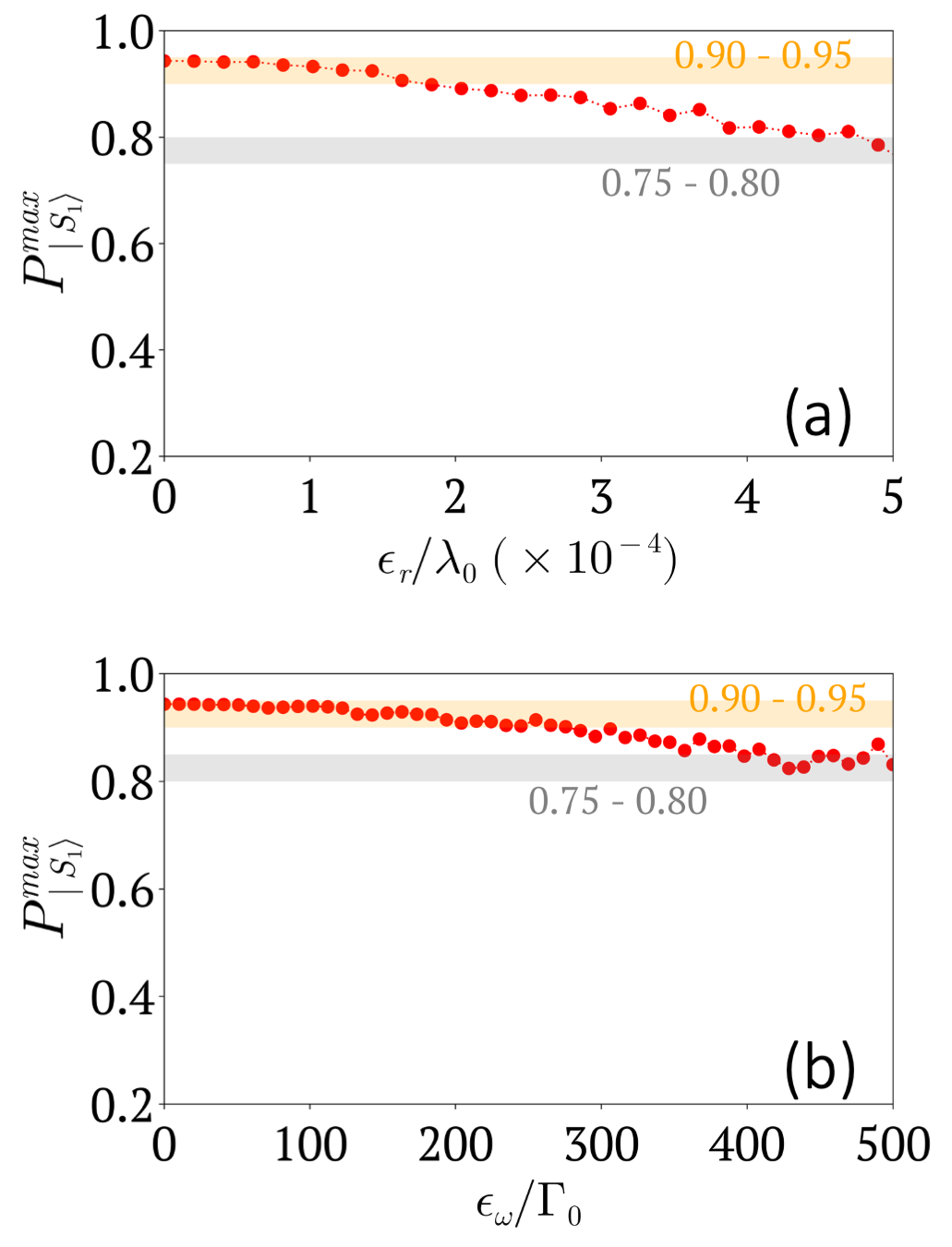}
\caption{Influence of scaled on-site position disorder $\epsilon_r/\lambda_0$ (a) and frequency disorder $\epsilon_{\omega}/\Gamma_0$ (b) on the maximum achievable population in state $\ket{S_1}$ ($P^{max}_{\ket{S_1}}$) via SUPER excitation. Over the displayed range, in both panels, $P^{max}_{\ket{S_1}}$ remains robust at 90--95\% (orange bar) and degrades to 75--80\% (gray bar) for the largest corresponding disorder values. The optimization parameters of Fig.~\ref{s1-inversion}(a) are utilized to populate state $\ket{S_1}$.}
\label{disorder-s1}
\end{figure}

Ref.~\cite{kerber:prr:2026} illustrates robustness of the population inversion mechanism to the collective states in the presence of (i) static position disorder, which may arise due to lattice vibrations, and (ii) on-site frequency inhomogeneities, which may originate from dephasing mechanisms (via substrate). In both cases, small perturbations, in particular on-site position displacements drawn from a normal distribution of width $\epsilon_r$ (in units of $\lambda_0$) or frequency inhomogeneities in which each emitter's transition frequency is drawn from a normal distribution centered at $\omega_0$ with standard deviation $\epsilon_\omega$ (in units of $\Gamma_0$), leave the pure bright and dark collective states largely unmixed. In our present case of the triangular trimer, the same physical picture remains valid. For minute position fluctuations, the dipole-dipole couplings $\Omega$ and the corresponding collective energy shifts remain nearly unchanged~\cite{kerber:prr:2026}. As a result, the dressed states retain their collective character. Similarly, reasonably weak on-site energy disorder induces only small mixing of the collective radiative states.

To quantify this robustness, we perform disorder-averaged simulations for both types of perturbation. For position disorder, we take a normal distribution of width $\epsilon_r$ (in units of $\lambda_0$) in all three spatial directions for each emitter, and average over 100 realizations for each disorder strength. The nearest-neighbor distance in our trimer is $d = 0.01\lambda_0$, hence the choice of the upper limit $\epsilon_r/\lambda_0 = 5 \times 10^{-4}$, which corresponds to $5\%$ of the inter-emitter spacing and ensures that the perturbed positions do not drastically alter the subwavelength geometry. Similarly, to account for frequency variations $\epsilon_\omega$ (in units of $\Gamma_0$), we consider a normal distribution with a standard deviation of $\epsilon_\omega$ for each emitter while keeping the positions fixed at the ordered trimer geometry, and average over 100 realizations. We consider frequency disorder strengths up to $\epsilon_\omega/\Gamma_0 \approx 500$, reflecting a moderate inhomogeneous broadening for instance in solid-state emitter platforms. In both cases, the system is evolved under the full master equation with the SUPER pulse sequences, and the maximum post-pulse population in the target collective radiative state is recorded. For the symmetric state $\ket{S_1}$, Fig.~\ref{disorder-s1} shows that $P^{max}_{\ket{S_1}}$ remains between 90--95\% over a broad range of both position (Fig.~\ref{disorder-s1}(a)) and frequency disorder (Fig.~\ref{disorder-s1}(b)), degrading to 75--80\% only for the largest disorder values considered. The antisymmetric state $\ket{B_1}$ exhibits comparable robustness (see Appendix~\ref{b1-pop}, Fig.~\ref{b1-inversion}(b,~c)): $P^{max}_{\ket{B_1}}$ stays within 85--95\% for moderate disorder and decreases to 70--80\% under strong position disorder and 66--76\% under strong frequency inhomogeneity. The slightly higher sensitivity of the prepared $\ket{B_1}$ compared to the case of $\ket{S_1}$ might be understood from the fact that the preparation of the antisymmetric state relies on precise site-dependent optical phases (for the symmetric state, all phases are zero), and thus appears to be more susceptible to both geometric and energetic perturbations. A recent study has shown that collective emission can survive inhomogeneous broadening even in subwavelength solid-state emitter arrays~\cite{bekenstein:arxiv:2026}. It is worth noting that the SUPER pulse operates on ultrashort (picosecond) timescales, and therefore accesses the collective states before the slower environmental fluctuations can strongly affect the system dynamics. In addition, the pulses induce an AC-Stark shift, which effectively decouples the system from the resonant and some near-resonant environmental modes. In this way, the `pure' electromagnetic layer of interaction becomes partially decoupled from the dissipative environment after the state preparation. This provides an additional physical reason why the population inversion mechanism via SUPER excitation remains robust at elevated temperature when compared to the conventional excitation scheme~\cite{Bracht:oq:2023}.

Therefore, for a deep-subwavelength triangular trimer, suitable pulse parameters can still lead to very high inversion efficiency into a chosen collective state, whether symmetric or antisymmetric, thereby allowing direct access to its distinct radiative properties~\cite{kerber:prr:2026}. The same idea may also be relevant for molecular aggregates and bio-inspired light-harvesting systems, where selective collective-state preparation could help isolate and probe the pure dipole-dipole interaction layer to directly witness their distinct radiative features. However, as discussed before, in biological originals the system-environment interaction is much more complex in reality, and a more accurate theoretical treatment would be required~\cite{pal:njp:2025}, which is currently beyond the scope of this manuscript.

\section{Experimental Platforms}
\label{expt}
Some experimental platforms may be very suitable for the efficient preparation of collective states in a deep-subwavelength triangular trimer of QEs with SUPER excitation, especially those relying on QDs and molecules. The possibility of fabricating very closely spaced semiconductor QDs~\cite{Dalacu_2019}, together with ongoing advances in strain-tunable quantum dots~\cite{trotta:prl:2018}, offers a hopeful view toward realizing such compact solid-state QE geometries in the future. In addition, DNA-origami-based assembly may provide an alternative platform for positioning QEs with nanometer-scale precision in tailored few-body configurations~\cite{liu:acsp:2018, Guillermo:acsn:2022}. Synthetic porphyrin rings~\cite{anderson:nc:2022, Kananenka:jpcc:2024} may also offer a useful molecular platform in this context. For optical addressing, the conventional implementation of the SUPER scheme, with sufficiently precise pulse shaping, parameter tunability~\cite{Karli:nl:2022}, and the adaptation of suitable near-field optical phase $\vartheta_j$ methods~\cite{kirchmair:np:2022, wallraff:prl:2010} (as applicable), should, in principle, be sufficient for selective state preparation~\cite{kerber:prr:2026} in such deep-subwavelength triangular trimer geometries. We note that for the symmetric state $\ket{S_1}$, where all optical phases vanish, a single global pulse should be sufficient. Selective preparation of states within the degenerate manifold, such as $\ket{B_1}$, requires site-dependent phase control, which at deep-subwavelength separations goes beyond far-field illumination and could be realized through near-field techniques or waveguide-mediated excitation~\cite{Dalacu_2019, tiranov:science:2023,
bach:prap:2025}. In practice, a reasonably controlled or weak system-environment interaction (which could, for instance, be obtained by choosing suitable platforms and/or through engineering suitable designs, as applicable) would help one achieve the expected outcomes.

\section{Conclusions and Outlook}
\label{conclusions}

Here we have investigated the selective preparation of specific collective excited states in a deep-subwavelength equilateral triangular trimer of two-level QEs using the SUPER scheme. We show that the preparation fidelity is governed by the spectral separation of the collective dressed states, depending crucially on the inter-emitter separation $d$. In the deep-subwavelength regime, i.e., $d = 0.01\lambda_0$, strong dipole-dipole interactions lead to very large collective energy shifts that well-resolve the dressed-state ladder and thereby enable near-unity inversion (approximately 94.3\%) into the symmetric collective state $|S_1\rangle$ and approximately 93.7\% into the antisymmetric state $\ket{B_1}$ (which is, in general, possible for any chosen collective state). As the inter-emitter separation increases to $d = 0.1\lambda_0$, the collective energy shifts diminish as the dressed states become nearly degenerate, the collective decay rates are modified, and the inversion efficiency drops substantially (to roughly 44\% in case of $\ket{S_1}$). Consequently, the final population gets distributed across multiple collective states (with the ground state also acquiring a substantial population); hence, the selective target state preparation scheme becomes significantly inefficient. This geometry- (or size-) dependent behavior clearly entails that the more compact the emitter arrangement is, the more favorable it is for SUPER-based selective collective-state preparation. Moreover, the degenerate manifolds introduced by the $C_3$ symmetry of the triangular geometry can be resolved through appropriate tuning of the site-dependent optical phases, demonstrating that the SUPER scheme offers state-selective control in the degenerate manifold distinct from the dimer case. In particular, very deep-subwavelength ($d\ll\lambda_0$) geometries of QEs are the most favorable candidates here. At this limit, the radiative character remains \textit{purer} over the characteristic timescale, which appears to allow for highly efficient inversion into a target state during this advanced pulsed excitation. Furthermore, the prepared collective states exhibit robustness against both static position disorder (with $P^{max}_{\ket{S_1}}$ remaining at 90--95\% over a broad range of disorder strength) and on-site frequency inhomogeneity, degrading to 75--80\% only at the largest disorder strengths considered (see Section~\ref{disorder}). The antisymmetric state $\ket{B_1}$ shows comparable resilience; however, its slightly higher sensitivity is perhaps due to its reliance on site-dependent optical phases during the excitation.

It is particularly encouraging given that the natural LH ring dimensions with nearest-neighbor separations of the order of $0.001\lambda_0$ lie in the regime where this excitation scheme could in principle perform seamlessly. However, it is important to note that the system-environment interaction is considerably more complex in biological systems. Within the partial applicability of the coupled-dipole model~\cite{pal:njp:2025}, eigenstates that lie closely spaced in the energy cascade, with correspondingly small collective energy shifts, are expected to remain hard to address selectively by this scheme. The robustness of the state preparation amid reasonable environmental decoherence (in terms of static position and frequency disorder) would inherit the very own essence of SUPER excitation mechanism~\cite{kerber:prr:2026}. Hence future experiments could help to isolate (up to a certain extent) and study the `pure' electromagnetic layer of interaction in LH complexes, and in doing so help establish the validity of quantum optics models in these systems. This would be a resourceful contribution to the evolving landscape of bio-inspired energy-efficient configurations~\cite{Mattiotti:njp:2021, Moreno-Cardoner:oe:22,erik:prxe:2023, pal:njp:2025, Erik:prl:2025}, offering an alternative and considerably simpler quantum optical description of these systems. This result will also be insightful for the development of appropriate coupled-dipole models to illustrate certain aspects relevant to quantum biology, such as a recent study in photosynthesis~\cite{pal:arxiv:2026}.

Looking ahead, we hope that our contribution motivates further exploration of pulse-based collective radiative state probing in bio-inspired and synthetic nanophotonic architectures. This will facilitate the study and observation of distinct radiative features, as well as their applicability in photonics and quantum information processing. The consistent observation that efficient preparation of collective states occurs at deep-subwavelength separations indicates that systems utilizing guided-mode-mediated interactions of QDs, for instance, fiber-mediated interactions of QDs~\cite{Dalacu_2019,tiranov:science:2023, bach:prap:2025}, might actually require very close inter-dot separations to achieve similar SUPER-excitation-based selective and highly efficient target radiative state preparation, which will be investigated more explicitly in the future.

Our work also points toward a timely intersection between collective quantum optics and Artificial Intelligence (AI)-assisted control. As these schemes scale to larger many-emitter architectures, the parameter space for pulse shaping and environmental engineering will quickly exceed the limits of manual optimization. Already for the trimer, the pulse optimization involves seven continuous parameters ($\alpha_1$, $\alpha_2$, $\Delta_1$, $\Delta_2$, $\sigma_1$, $\sigma_2$, $\tau$) together with the discrete choice of site-dependent optical phases $\vartheta_j$, and this complexity will grow rapidly with the number of QEs. Looking forward, AI-assisted techniques, reinforcement-learning-based searches~\cite{briegel:sr:2012, Schuff:njp:2020}, neural-network-based adaptive methods~\cite{fiderer:prxq:2021}, offer a promising avenue to efficiently discover optimal pulse sequences for targeted radiative state preparation. Establishing such an optimized open-source framework represents a natural next direction to pursue.

Furthermore, the triangular trimer, or more generally compact subwavelength ring geometries, could support darker subradiant states~\cite{Ostermann:oe:12, zhu:pra:2013,moreno:pra:2019} than other subwavelength QE arrangements. The SUPER-excitation-based preparation of both the symmetric superradiant and antisymmetric subradiant collective states demonstrated in this work, together with their robustness to reasonable static positional and frequency imperfections, could hold great promise for quantum metrology, as recently suggested in the contexts of subradiant precision sensing~\cite{yelin:pra:2026} and superradiance-based magnetometry under dephasing~\cite{sakuma2026arxiv}. In particular, the ability to deterministically prepare either a superradiant or a subradiant state in the same geometry, simply by tuning the site-selective optical phases, could enable direct comparative measurements of their respective metrological advantages within a single experimental platform. The superradiant state, with its enhanced collective decay rate, could improve signal strength per measurement cycle, while the subradiant state, with its suppressed decay, could extend coherence times and thereby enhance frequency precision. This resilience of the prepared collective states, whether bright or dark, seems to align with recent studies demonstrating that correlated-state protocols can retain a certain advantage even under realistic decoherence and noise conditions~\cite{hammerer:sa:2024, Kielinski:rpp:2025}. A detailed quantification of these metrological advantages is reserved for future study.

\section{Data Availability}
The simulations in this manuscript were performed using the QuantumOptics.jl~\cite{KRAMER:cpc:2018} and CollectiveSpins.jl~\cite{collectivespins} frameworks in the Julia programming language. The panels were prepared using the Julia framework library~\cite{Danisch2021}. The data that support the findings of this manuscript are openly available at~\cite{zenodo:nunner}.

\section*{Acknowledgements}
A.P. gratefully acknowledges her participation in the `NanoLight 2026' (Centro de Ciencias de Benasque Pedro Pascual), `Artificial Intelligence Photonics 2026' (San Sebastián), and the `57th Conference of the European Group on Atomic Systems' (Toruń); the insights gained during these visits inspired and helped to craft some discussions presented in this manuscript. This research was funded in whole or in part by the Austrian Science Fund (FWF) 10.55776/ESP682. We also gratefully acknowledge support from FWF projects: 10.55776/COE1 (quantA) and 10.55776/FG5 (Forschungsgruppe FG 5).

\begin{figure}[b]
\centering
\includegraphics[width=\linewidth]{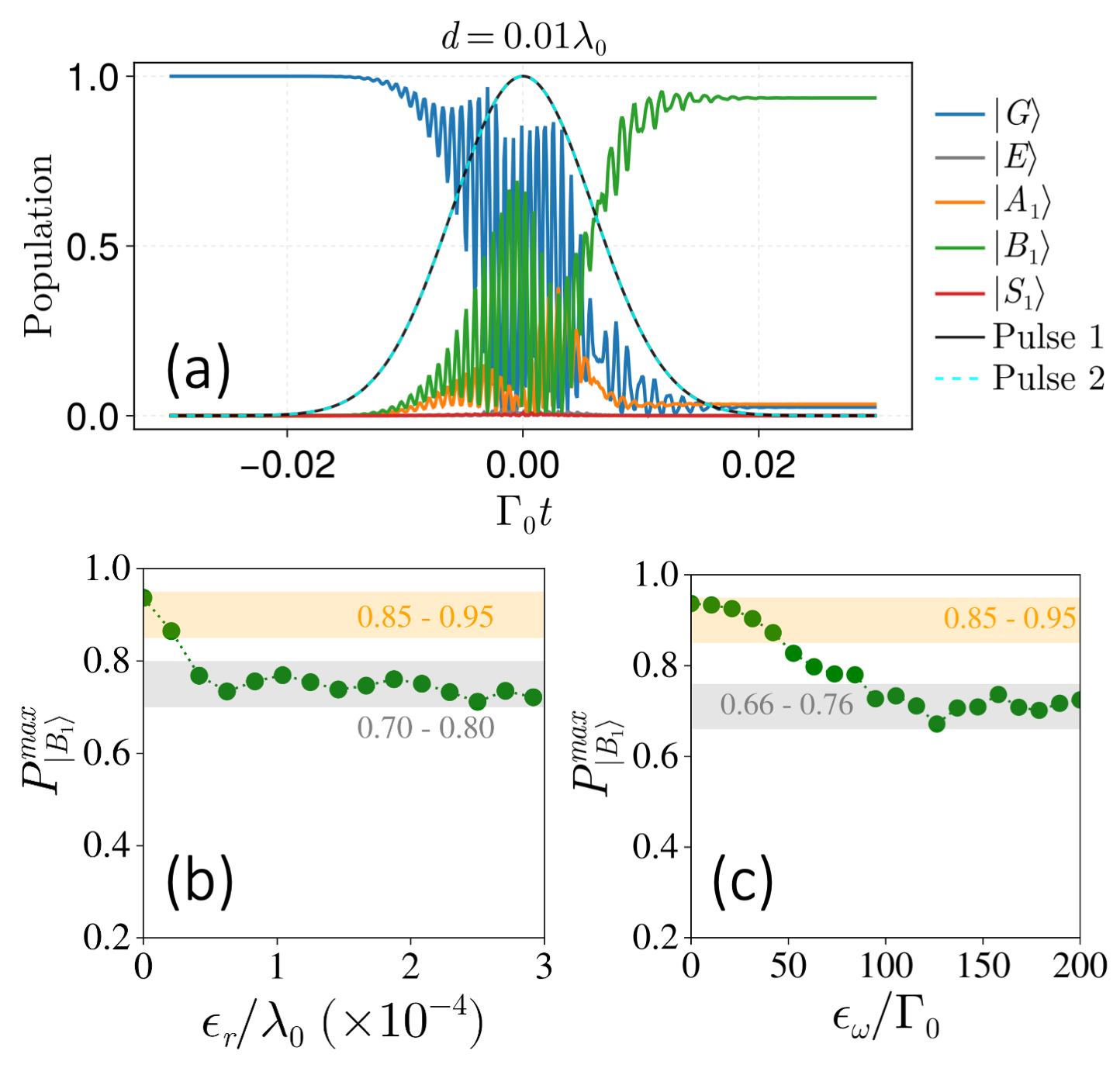}
\caption{Panel (a) shows the inversion into the antisymmetric collective state $\ket{B_1}$ for dipoles oriented transversely to the plane of the triangle. The inversion reaches approximately 93.7\% using SUPER pulses with parameters $(\alpha_1,\alpha_2)=(28\pi,12\pi)$, $\tau=0$ ps, $(\Delta_1,\Delta_2)=(-5.4,-10.6)$ meV, $(\sigma_1,\sigma_2)=(6,6)$ ps, and optical phases $\vartheta_1=1.8\pi$, $\vartheta_2=0.2\pi$, and $\vartheta_3=0.9\pi$. Panels (b) and (c) show the maximum population in $\ket{B_1}$, i.e., $P^{\max}_{\ket{B_1}}$, as a function of on-site position disorder $\epsilon_r/\lambda_0$ and frequency inhomogeneity $\epsilon_{\omega}/\Gamma_0$, respectively. Over the displayed range, $P^{\max}_{\ket{B_1}}$ remains in the 85--95\% range (orange) and decreases to 70--80\% (gray) in (b) and 66--76\% in (c) at the largest disorder values. The optimization parameters from panel (a) are used throughout.}
\label{b1-inversion}
\end{figure}
\appendix
\section{Coherent and Dissipative Dipole-Dipole Couplings}
\label{green}
The Green-tensor $\boldsymbol{G}(\boldsymbol{r},\omega_0)$ in free-space reads as follows
\begin{align}
\boldsymbol{G}(\boldsymbol{r},\omega_0) &= \frac{e^{ik_0 r}}{4\pi k^2_0 r^3}\bigg[\bigg(k^2_0r^2 +i k_0 r-1\bigg)\boldsymbol{1} \nonumber\\
    &- \bigg(k^2_0r^2 + 3i k_0r-3\bigg)\frac{\boldsymbol{r}\otimes\boldsymbol{r}}{r^2}\bigg]~,
\end{align}
with the wave number denoted by $k_0 = \omega_0/c$ corresponding to the atomic transition frequency $\omega_0$ and $r = |\boldsymbol{r}|$. The energy shift $\Omega_{ij}$ (coherent dipole-dipole coupling) can be written as
\begin{align}
    \Omega_{ij} = -\frac{3\pi\Gamma_0}{k_0}\hat{\boldsymbol{\mu}}_i\cdot\mathrm{Re}\{\boldsymbol{G}(\boldsymbol{r}_{ij},\omega_0)\}\cdot\hat{\boldsymbol{\mu}}_j~,
\end{align}
where $\Gamma_0$ is the single emitter spontaneous decay rate and $\hat{\boldsymbol{\mu}}_i, \hat{\boldsymbol{\mu}}_j$ are the unit vectors of the corresponding dipole moments of $i^{th}, j^{th}$ QE, respectively. Using the position vectors $\boldsymbol{r}_{i}$ and $\boldsymbol{r}_{j}$, the distance vector is determined via $\boldsymbol{r}_{ij} = \boldsymbol{r}_{i} - \boldsymbol{r}_{j}$. The collective decay rates $\Gamma_{ij}$  (dissipative dipole-dipole coupling) can be expressed as
\begin{align}
    \Gamma_{ij} = \frac{6\pi\Gamma_0}{k_0}\hat{\boldsymbol{\mu}}_i\cdot\mathrm{Im}\{\boldsymbol{G}(\boldsymbol{r}_{ij},\omega_0)\}\cdot\hat{\boldsymbol{\mu}}_j~.
\end{align}
Both $\Omega_{ij}$ and $\Gamma_{ij}$ are distance and dipole orientation dependent quantities~\cite{FICEK:PR:2002}.

\section{Selective Preparation of the Antisymmetric Collective State}
\label{b1-pop}
Here we show the efficient preparation of the antisymmetric state $\ket{B_1}$ (see Table~\ref{tab:eigen}) in the same deep-subwavelength trimer ($d = 0.01\lambda_0$), using similar choices of SUPER pulse envelopes employed earlier but with a different choice of relative optical phases $\vartheta_j$. As discussed in the main text, $\ket{B_1}$ and $\ket{A_1}$ are energetically degenerate (see Fig.~\ref{schematic}(b), Table~\ref{tab:eigen}) due to $C_3$ symmetry. To selectively target $\ket{B_1}$, the optical phases $\vartheta_j$ are first chosen carefully; the remaining pulse parameters are then optimized to maximize the inversion efficiency. Fig.~\ref{b1-inversion}(a) shows that suitable pulse parameters, together with appropriate choices of optical phases $\vartheta_j$~\cite{kerber:prr:2026}, yield approximately 93.7\% inversion into $\ket{B_1}$. This confirms that the site-dependent optical phases provide sufficient control to resolve the degenerate manifold and selectively prepare a chosen state within it. For larger inter-emitter separations, this efficiency decreases, as detailed in the main text. Fig.~\ref{b1-inversion}(b) and (c) show the robustness of the achieved population against on-site position disorder $\epsilon_r/\lambda_0$ and frequency inhomogeneity $\epsilon_{\omega}/\Gamma_0$, respectively. $P^{\max}_{\ket{B_1}}$ remains within 85--95\% for moderate disorder, decreasing to 70--80\% (position) and 66--76\% (frequency) at the largest disorder values. The slightly higher sensitivity of $\ket{B_1}$ compared to $\ket{S_1}$ (cf.\ Fig.~\ref{disorder-s1}) is likely due to its reliance on precise site-dependent optical phases. A systematic investigation of this phase-profile dependence in the presence of disorder, potentially aided by automated optimization techniques (see Section~\ref{conclusions}), represents a future avenue to trace.

\bibliography{ref}

\end{document}